# Dependency of high-speed write properties on external magnetic field in spin-orbit torque in-plane magnetoresistance devices

Yohei Shiokawa[1], Eiji Komura[1], Yugo Ishitani[1], Atsushi Tsumita[1], Keita Suda[1], Kosuke Hamanaka[1], Tomohiro Taniguchi[2], and Tomoyuki Sasaki[1]

*[1]TDK Corporation, Advanced Products Development Center, Ichikawa, Chiba 272-8558, Japan*

*[2]National Institute of Advanced Industrial Science and Technology (AIST), Research Center for Emerging Computing Technologies, Tsukuba, Ibaraki 305-8568, Japan*

E-mail: yshiokawa@jp.tdk.com

**Abstract**

Spin-orbit torque (SOT) magnetoresistance (MR) devices have attracted attention for use in next-generation MR devices. The SOT devices are known to exhibit different write properties based on the relative angle between the magnetization direction of the free layer and the write-current direction. However, few studies that compare the write properties of each type have been reported. In this study, we measured the external perpendicular-magnetic field dependence of the threshold write current density and the write current switching probability using two types of in-plane magnetization SOT-MR devices.

**Text**

Spin-orbit torque (SOT) magnetoresistance (MR) devices have attracted attention owing to their high speed, high endurance, and low power consumption. A high-speed write operation spanning less than 0.5 ns has already been conducted.[1-6)] Because the in-plane write current only passes through the SOT wiring, the barrier layer in the magnetic tunnel junction (MTJ) can prevent breakdown. The write endurance of the three-terminal SOT device is expected to be higher than that of the two-terminal spin-transfer torque (STT) device. Experimental write endurance of up to $10^{12}$ cycles with tungsten (W)-SOT wiring has been reported.[7] The SOT device can decrease the write current with size scaling of the narrow SOT wiring because, the SOT wiring as β-W is made of metal.[8]

The SOT device is known to exhibit different write properties based on the relative angle between the magnetization direction of the free layer and the write current direction.[9] The three types of representative examples are as follows: the first has a perpendicular magnetization direction called type-Z;[1,5,6,10-14] the second has an in-plane magnetization direction parallel to the write current, called type-X;[9,15] the third has an in-plane magnetization direction orthogonal to the write current, called type-Y.[7,16,17] The type-Y device can be switched with no external magnetic field because the direction of the polarized spin current induced by the write current in the SOT wiring is anti-parallel to the magnetization direction of the free layer, as an anti-damping torque. Conversely, type-Z and type-X devices require an external magnetic field because the direction of spin polarization is orthogonal to the magnetization direction of the free layer. However, it is expected that this orthogonally polarized spin current can realize high-speed magnetization switching with a lower threshold write current density $J_{th}$ than a type-Y device.[9] However, few studies have experimentally investigated the switching properties of type-X devices.[15]

In this study, we measured the dependence of $J_{th}$ and write current switching probability on an external perpendicular magnetic field $H_z$ using type-Y and type-X devices with the same in-plane magnetization type. We compared the write current and switching properties of type-Y and type-X devices. Moreover, because the required external perpendicular magnetic field in a type-X device may depend on the effective in-plane magnetic anisotropy field,[18] we estimated the effective in-plane magnetic anisotropy field using spin-torque ferromagnetic resonance (ST-FMR) measurements.[19-21]

The type-Y and type-X devices used the same MTJ film structure. The MTJ used in this study was a top-pinned type MTJ with a tungsten layer as the SOT wiring. The stacking layer included a thermal oxide Si substrate/W (3 nm)/CoFeB (2 nm)/MgO (1.8 nm)/CoFeB (2 nm)/Ru (0.8 nm)/CoFe (2 nm)/IrMn (8.5nm)/Ru (2 nm)/Ta (3 nm) Post-annealing was conducted at 250 °C for 3 h in an in-plane magnetic field of over 1 T. The resistivity of the W-SOT was 233 μΩ cm. We fabricated both type-Y and type-X devices using this MTJ film by applying photolithography and an ion milling process. Figure 1 (a) shows a schematic illustration of the type-Y device. The W-SOT wiring was 3-nm thick, 700-nm long, and 420-nm wide; its shape was elliptical with a 297-nm long axis and 122-nm short axis. The type-Y device had a long axis of MTJ orthogonal to the direction of the write current, and the type-X device had a long axis of MTJ parallel to the direction of the write current. Figures 1 (b) and 1(c) show the R–H curves of type-Y and type-X devices in $H_x$, $H_y$, and $H_z$. In the type-Y device, the MR ratio was 84.0% and $R_{min}$ was 32.3 kΩ while the type-X device MR ratio was 76.7%, and the $R_{min}$ was 31.5 kΩ. The saturation field of the in-plane hard axis was approximately 500 Oe for both types. We fabricated MTJs of approximately the same level between the type-Y and type-X devices. Moreover, the thermal stability Δ of the type-Y device was 58.3 and measured using the switching probability of the external pulse magnetic field (pulse width 1 s)[21,22] (Figure 1 (d)).

First, we measured the $J_{th}$ of the type-Y and type-X devices. Figure 2 (a) shows the relationship between the $J_{th}$ and the varied pulse width in the type-Y and type-X devices.

We note that the $J_{th}$ of the type-X device depends on the external perpendicular magnetic field. Under each of the pulse width and $H_z$ conditions, we measured the five R–J curves and calculated the average value of $J_{th}$ by five of $+J_{th}$ from the parallel state (P) to the anti-parallel state (AP) and five of $-J_{th}$ from AP to P.[7] The type-Y device used no external magnetic field and the type-X device used $H_z$ ranging from 600 to 1600 Oe. In both cases, $J_{th}$ increased as the pulse width decreased to less than 10 ns. However, the increasing trend of $J_{th}$ with the decreasing pulse width in the type-X device was lower than that in the type-Y device consistently with the numerical simulations.[9] By comparing the absolute values of $J_{th}$ between the type-Y device and type-X device, the $J_{th}$ of the type-X device for a pulse width of 1 ns at 600 Oe was higher than that of the type-Y device. Figure 2 (b) shows a plot of $J_{th}$ as a function of $H_z$ at pulse widths of 1 ns, 2 ns, 5 ns, and 10 ns. In the high-speed write pulse region of 1 ns, the type-X device required $H_z$ higher than 800 Oe, for the $J_{th}$ of the type-X device to be lower than that of the type-Y device.

Next, we discuss the write current switching probability. Plots of the switching probability within 100 write cycles as a function of the write current density and the normalized current density are shown in Figure 3. The pulse width was varied for 10 ns, 5 ns, and 1 ns. The type-Y device was measured under zero external magnetic fields and the type-X device was measured with $H_z$ as 600 Oe, 800 Oe, and 1000 Oe. To compare the gradient of the write current switching probability, the lines in Figure 3 (d)–(f) serve as guides to the eye based on Equation (1) as follows:[22,23]

$$P = 1 - \exp\left[-\frac{\tau_P}{\tau_0}\exp\left\{-\Delta\left(1 - \frac{J}{J_c}\right)\right\}\right] \qquad (1)$$

$\tau_P$ is the pulse width, $\tau_0$ is the switching speed of magnetization (assumed 1 ns), $\Delta$ is the thermal stability, $J$ is the applied write current density, and $J_c$ is the critical current density as the fitting parameter. Whereas Eq. (1) has been used for two-terminal MR devices as well as type-Y SOT-MR devices, we note that the applicability to type-X devices has not been studied yet. The gradient of the write current switching probability against the $J/J_c$ of the type-Y device remains unchanged from 10 ns to 1 ns. In contrast, the gradient of the write current switching probability of the type-X device shows a precipitous change with increasing $H_z$ and decreasing pulse width. It is suggested that the type-X device requires a precisive external magnetic application system and a precisive pulse-generator write circuit because the variations of the perpendicular magnetic field $H_z$ and pulse width involve the fluctuation of probability.

According to these results, the type-X device showed a lower $J_{th}$ and precipitous write probability than the type-Y device in the presence of a high perpendicular magnetic field $H_z$. A similar external magnetic-field dependence in a type-Z device is predicted.[12,14,23)] Although some studies have proposed a method to apply a magnetic field to a device with a permanent magnet,[2,6)] there are some concerns regarding the generation of high intensity external magnetic fields, the stability of the permanent magnet, the effect of a stray field on a neighbor element, and so on. Because there is no need for an external magnetic field for switching, a type-Y device might be appropriate for SOT-MRAM.

Finally, the required intensity of $H_z$ in the type-X device for SOT switching is discussed. According to Ref. 18, the threshold write current density $J_{th}$ and the required intensity of $H_z$ of the type-X device depend on the in-plane magnetic anisotropy field $H_{kin}$ and the demagnetization field $H_{demag}$. In this study, the $H_{kin}$ and the $H_{demag}$ were estimated using the ST-FMR measurement induced by SOT. We used a type-Y device for the ST-FMR measurement because it was expected to display a large signal compared with the type-X device. Figure 4 (a) shows a schematic illustration of the ST-FMR measurement. In the type-Y device element, a -10 dBm AC signal was applied to the SOT wiring using a signal generator operating at 1–5 GHz, and the output voltage of the MTJ was measured using a lock-in amplifier. The external magnetic field was in the z-direction $(H_z)$, ranging from 400 to 3000 Oe. The dependence of $f$, defined as the first peak at $H_z$, is illustrated in Figure 4 (b). This experimental relation of the first peak was adapted to Kittel's FMR equation,[24] which is expressed as follows:

$$f = \frac{\gamma}{2\pi}\sqrt{H_{kin}\left(\frac{-H_z^2}{H_{demag}+H_{kin}-H_{k\perp}}+H_{demag}+H_{kin}-H_{k\perp}\right)} \quad (3)$$

$H_{kin} = 4\pi M(N_y - N_x)$

$H_{demag} = 4\pi M(N_z - N_y)$

where $H_{kin}$ is the in-plane magnetic anisotropy field caused by shape anisotropy, $H_{k\perp}$ is the perpendicular magnetic anisotropy field, and $H_{demag}$ is the demagnetization field. The demagnetization factors were calculated as $N_z = 0.9595$, $N_x = 0.0089$, and $N_y = 0.0315$ based on the ellipse MTJ of a = 297 nm, b = 122 nm, and t = 2 nm.[25] The fitting result to the first peak was as follows: M = 1540 emu/cc, $H_{kin}$ = 440 Oe, $H_{demag}$ = 17.9 kOe, and $H_{k\perp}$ = 14.4 kOe. The effective in-plane magnetic anisotropy field was determined as $H_{demag} + H_{kin} - H_{k\perp} = 3.9$ kOe. It was found that the free layers of our prepared type-Y and type-X device elements not only exhibited in-plane magnetization, but also high perpendicular anisotropy. This perpendicular anisotropy may have been caused by both the interface magnetic anisotropy of CoFeB/MgO[26,27] and W/CoFeB.[28-30] The increase in the interface magnetic anisotropy of the free layer could decrease the required perpendicular magnetic field $H_z$ in the type-X device.

To verify this argument, we perform numerical simulations of the Landau-Lifshitz-Gilbert (LLG) equation at finite temperature; see also supplementary data for details of the calculations. Figure 5 (a) shows a histogram of the switching current densities for the type-X device obtained by the LLG simulation, where the external magnetic field $H_z$ and the pulse width are 600 Oe and 5 ns, respectively. The averaged switching current density,

$0.71 \times 10^8$ A/cm$^2$, is almost same with the experimental result shown in Figure 2 (a). Note that $H_{k\perp}$ = 14.4 kOe used in the LLG simulation is derived from the experimental result mentioned above. Figures 4 (b) – 4 (d) show the histograms of the switching current densities, where $H_z$ is (b) 400, (c) 200, and (d) 100 Oe. The averaged values of the switching current density in these figures, (b) 0.70, (c) 0.70, and (d) 0.69 $\times 10^8$ A/cm$^2$, are almost same with that found in Figure 4(a). The values of the perpendicular magnetic field $H_{k\perp}$ used in the simulations are (b) 15.7, (c) 16.8, and (d) 17.4 kOe, respectively. These results indicate that the value of the perpendicular magnetic field $H_z$ can be decreased by increasing the perpendicular magnetic anisotropy. It should, however, be noted that the reduction of the external magnetic field is much smaller than the enhancement of the perpendicular magnetic anisotropy field. For example, to reduce the external magnetic field from 600 to 400 Oe, the perpendicular magnetic anisotropy field should be increased 1.3 kOe. In this sense, we consider that an enhancement of the perpendicular magnetic anisotropy is inefficient to reduce the power consumption of type-X device. In addition, a finite magnetic field is necessary for type-X device to achieve a deterministic switching in principle, even if we can achieve a high perpendicular magnetic anisotropy.

In conclusion, we measured the external perpendicular magnetic field dependences of $J_{th}$ and the write current switching probabilities in type-X and type-Y devices. The dependence of $J_{th}$ on the pulse width in type-X device is weak compared with that in the type-Y device. In addition, switching probability in type-X device shows more precipitous change with respect to the current than that of the type-Y device. However, the type-X device required a high external magnetic field for switching. For example, in the high-speed write pulse region of 1 ns, an $H_z$ higher than 800 Oe is required in the type-X device to make $J_{th}$ lower than that in the type-Y device. Although the LLG simulation shows that the required $H_z$ can be reduced by increasing the perpendicular magnetic anisotropy, it is inefficient because the reduction is smaller than the required enhancement of the magnetic anisotropy field. In addition, in practical devices, type-X device will face several issues, such as the necessary of a finite magnetic field for the deterministic switching, the stability of the magnetic field, and the effect of a stray field on a neighbor element. Therefore, we consider that type-Y device will be useful for SOT-MR devices.

**Acknowledgments**

The authors acknowledge Masamitsu Hayashi, Shinji Isogami, and Seiji Mitani for participating in valuable discussions.

and W. Zhao, IEEE Trans. Magn. **54**, 1300705 (2018)

**Figure Captions**

**Fig. 1.** (a) Schematic illustration of type-Y element; (b) *R-H* curves of type-Y and (c) *R-H* curves of type-X (external magnetic field along the x-axis ($H_x$), y-axis ($H_y$), and z-axis ($H_z$)); (d) switching probability of external pulsed magnetic field with type-Y, obtained thermal stability $\Delta$ = 58.3.

**Fig. 2.** (a) Relationship of $J_{th}$ and the varying pulse width in type-Y and type-X devices, depending on $H_z$ (b) plot of $J_{th}$ as a function of $H_z$ at pulse widths of 1 ns, 2 ns, 5 ns, and

10 ns with type-Y ($H_z = 0$ Oe) and type-X devices.

**Fig. 3.** Write current switching probability in 100 write cycles as a function of the write current density and normalized current density for type-Y and type-X devices, with pulse widths of (a), (d) 10 ns, (b), (e) 5 ns, and (c), (f) 1 ns. Note that (a), (c), and (e) are identical to those in (b), (d), and (f), except the fact that the current densities in (b), (d), and (f) are normalized by $J_c$.

**Fig. 4.** (a) Schematic illustration of the measurement system for ST-FMR with type-Y; –10 dBM applied AC signal via SOT wiring using signal generator providing 1–5 GHz (b) dependence of $f$, defined as the first peak on $H_z$; (the solid line was the fitting line from Kittel's FMR equation).

**Fig. 5.** Histograms of the switching current density obtained by the LLG simulations at finite temperature for 1000 events, where the perpendicular magnetic fields are (a) 600, (b) 400, (c) 200, and (d) 100 Oe. The averaged switching current densities are (a) 0.71, (b) 0.70, (c) 0.70, and (d) $0.69 \times 10^8$ A/cm$^2$. The interface perpendicular magnetic anisotropy fields $H_{k\perp}$ are (a) 14.4, (b) 15.7, (c) 16.8, and (d) 17.4 kOe.

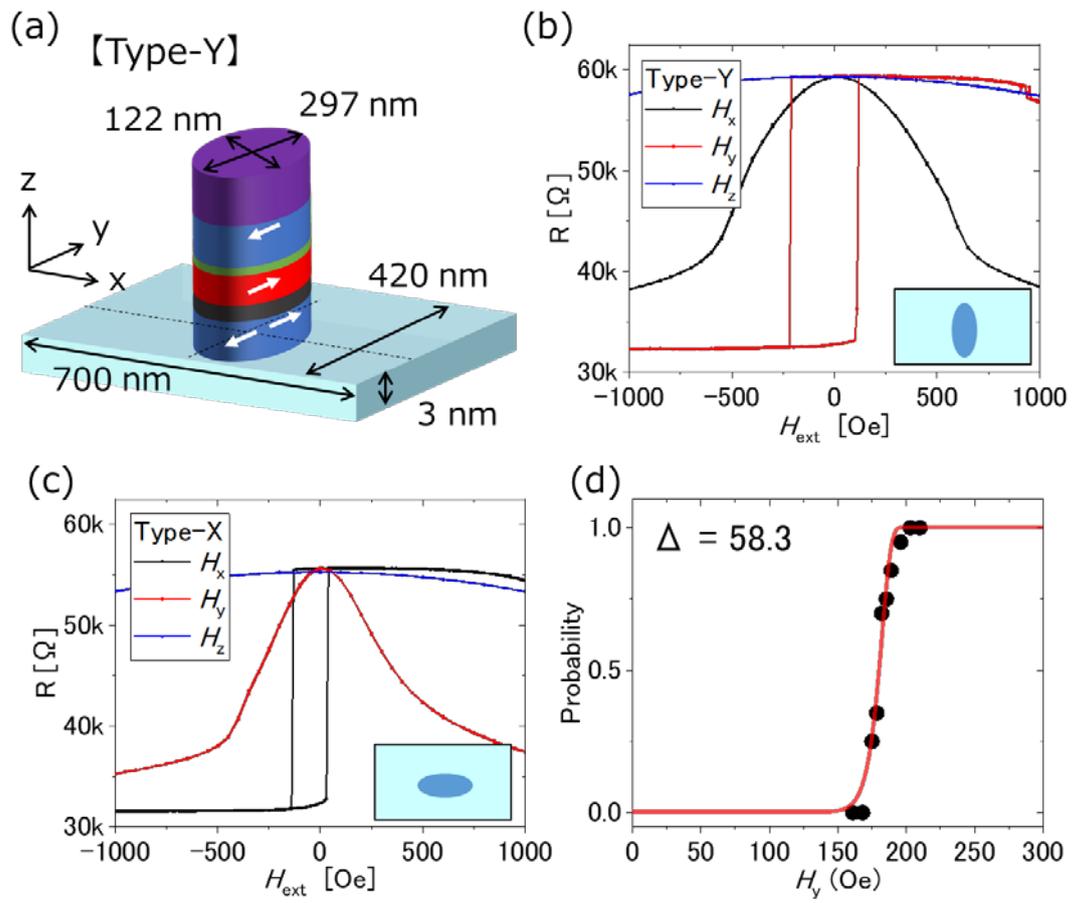

Fig. 1.

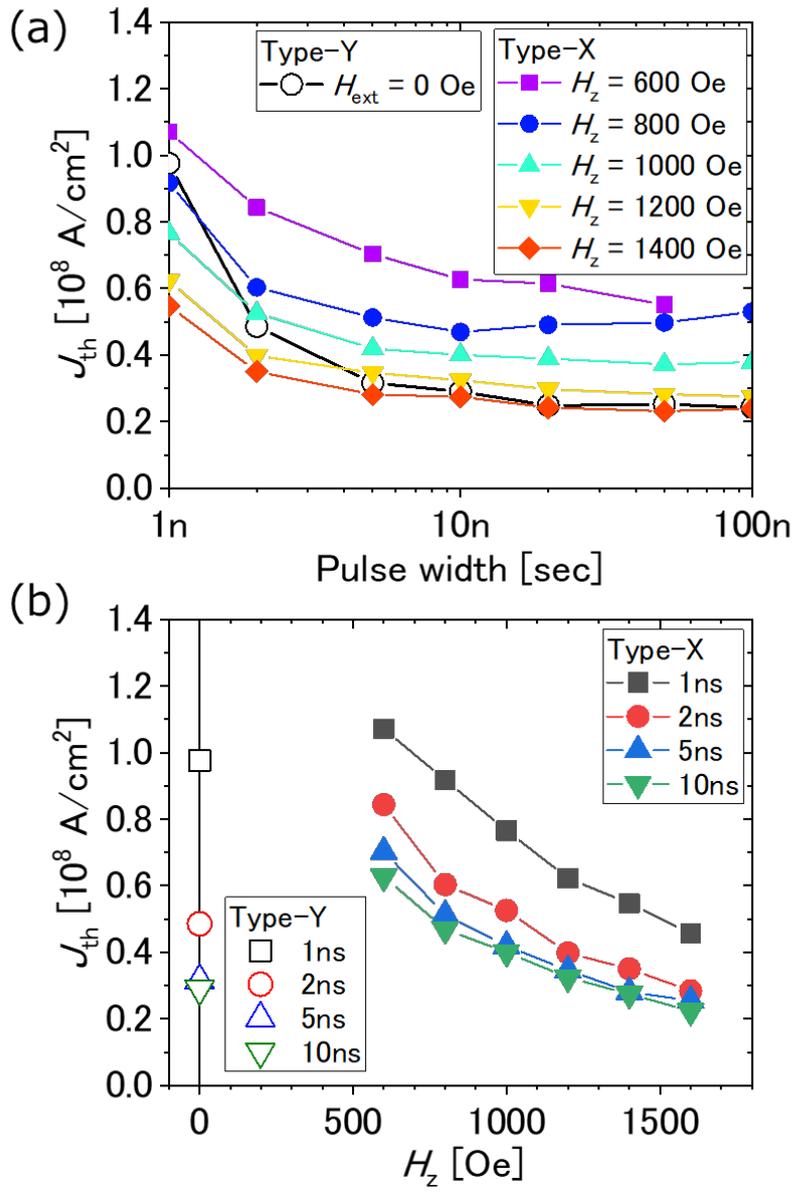

Fig. 2.

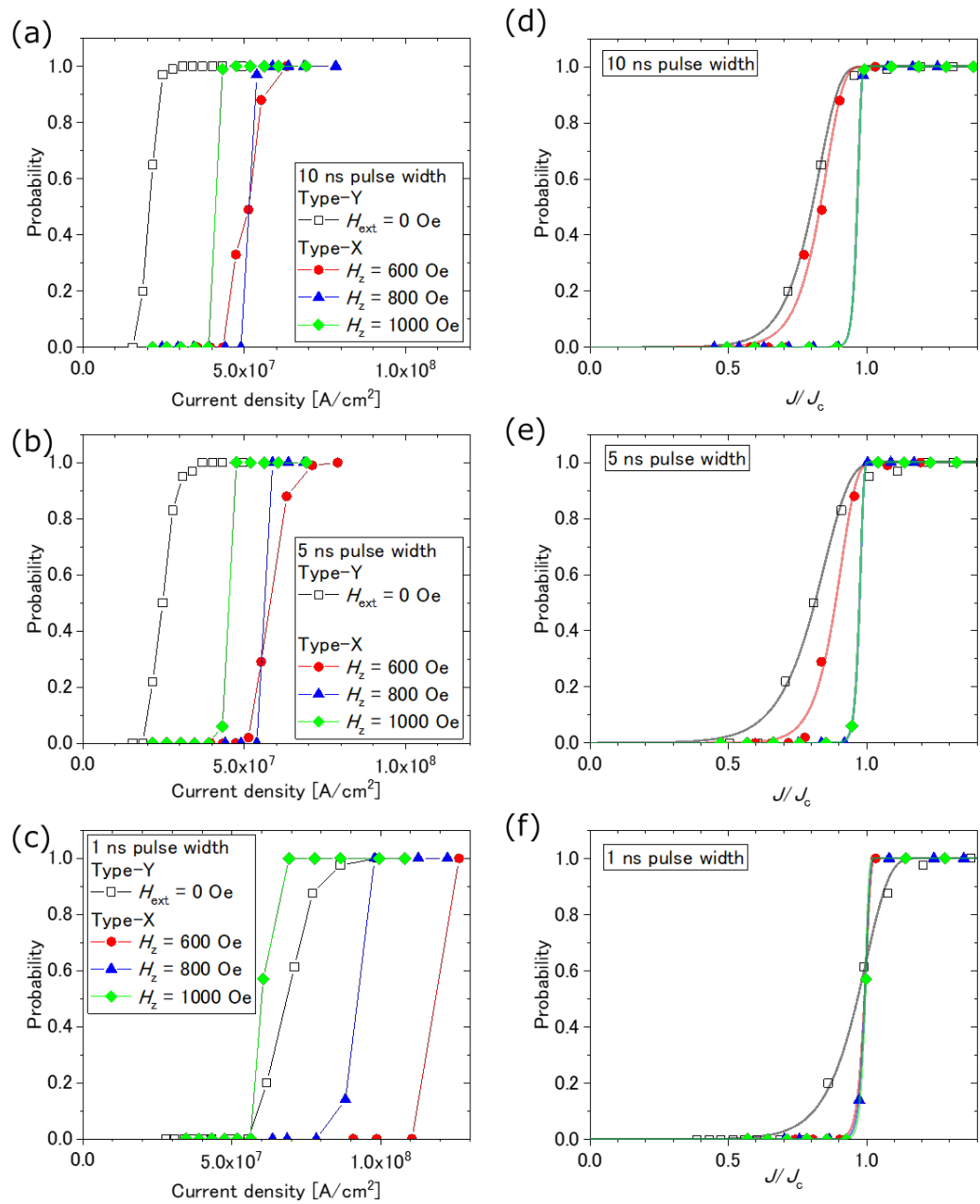

Fig. 3.

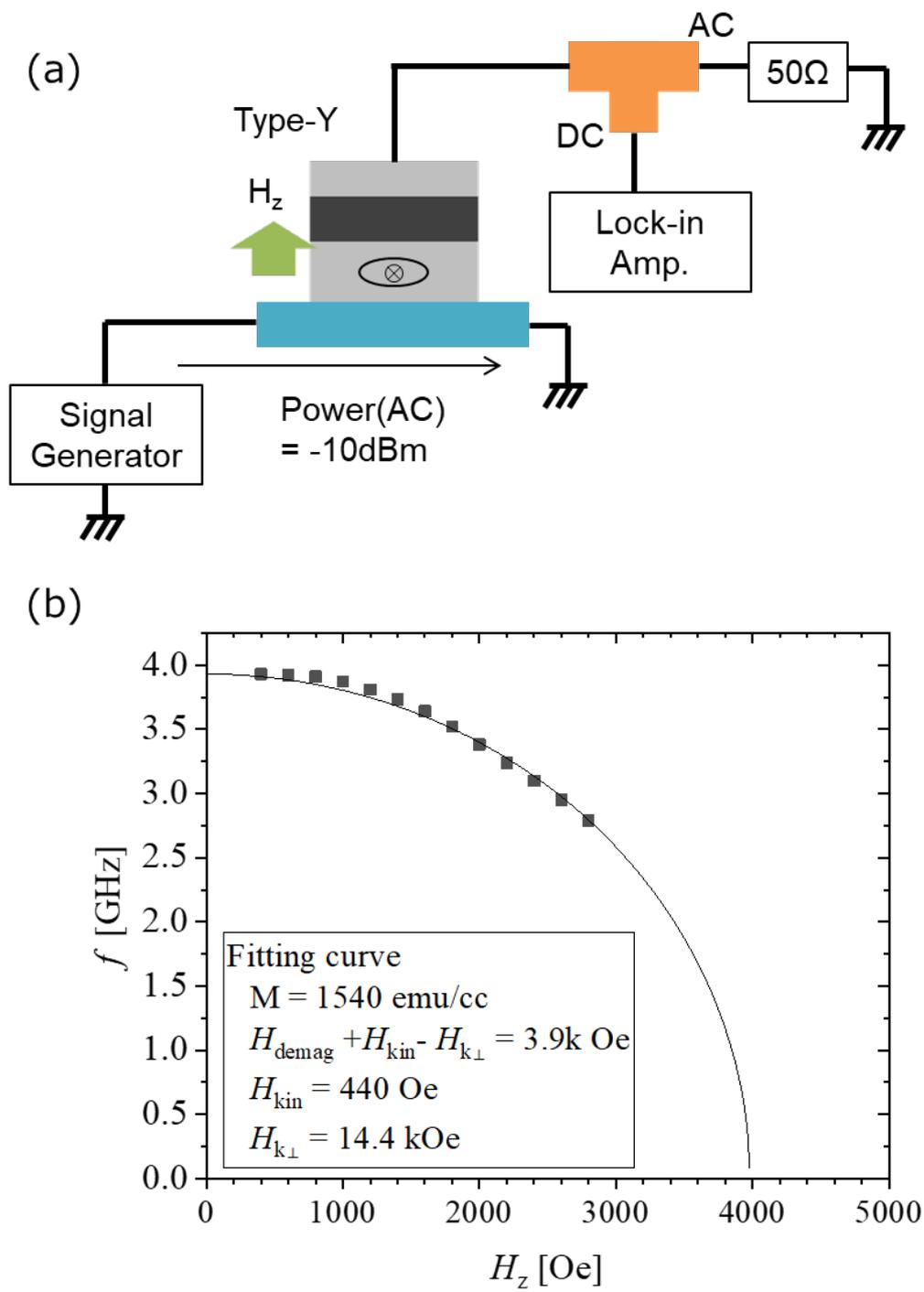

Fig. 4.

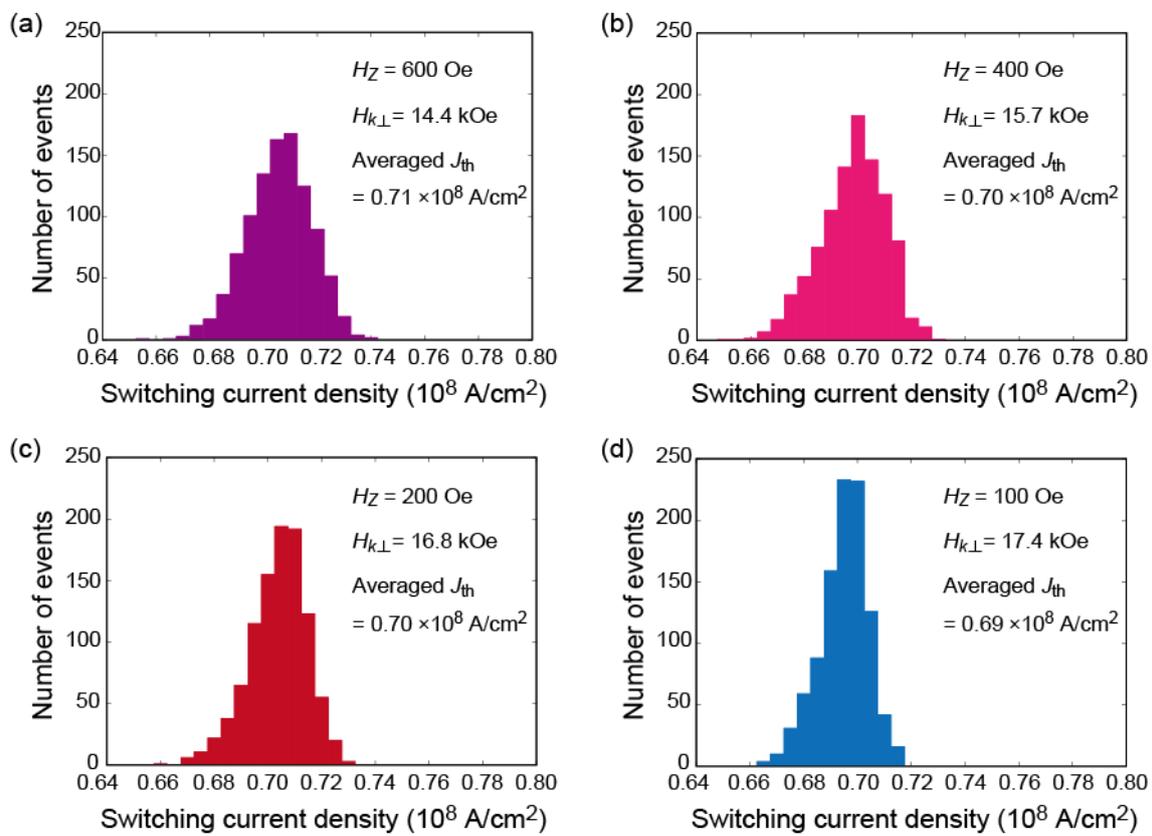

Fig. 5.